# A Miniature Potentiostat for Impedance Spectroscopy and Cyclic Voltammetry in Wearable Sensor Integration

*Franci Franulovic[1], Shawana Tabassum[1,#]*

[1]*Department of Electrical and Computer Engineering, The University of Texas, Tyler; Texas, U.S.A.*

[#]*Corresponding Author: Shawana Tabassum: stabassum@uttyler.edu*

# Abstract

A potentiostat is an analytical device and a crucial component in electrochemical instruments used for studying chemical reaction mechanisms, with potential applications in early diagnosis of disease or critical health conditions. Conventional potentiostats are typically benchtop devices designed for laboratory use, whereas a wearable potentiostat can be interfaced with biochemical sensors for disease diagnostics at home. This work presents a low-power potentiostat designed to connect with a sensor array consisting of eight to ten working electrodes. The potentiostat is capable of running Electrochemical Impedance Spectroscopy and Cyclic Voltammetry. The system is powered by lithium-ion batteries and uses Bluetooth for data transmission to the user. A single ARM M4 microcontroller, integrated with a Bluetooth low-energy radio module (Silicon Labs EFR32BG13 SoC), controls the entire system. The potentiostat's accuracy, reliability, and power efficiency were evaluated and compared against existing commercial benchtop potentiostats. Additionally, we have outlined future steps to enhance circuit miniaturization and power efficiency, aiming to develop fully integrated wearable sensing devices comparable in size to a wristwatch.

# Introduction

The COVID-19 pandemic underscored the need for affordable and rapid testing, as timely detection, such as for COVID-19, helps prevent disease spread. Unfortunately, many health conditions go undetected until symptoms appear, leading to more complex and costly treatments that could have been avoided with early detection. Early detection is a challenging task requiring various screening techniques that are constantly being refined for greater effectiveness. For over 40 years, researchers have sought new solutions based on their findings[1]. Electronic sensing devices, like Electrocardiograms (ECG) and pulse oximeters, have been used in medicine for decades. However, there are currently no wearable electrochemical instruments for disease detection at home. Advances in the semiconductor and communication industries now enable the design of compact, battery-powered devices capable of processing and analyzing data from wearable sensors at the edge. These sensors monitor physiological and biochemical markers, with numerous potential applications in areas such as remote health monitoring[2–11] and agricultural crop health monitoring[12–18].

Wearable sensors should not interfere with the wearer's movement or daily activities. Therefore, they must be small and flexible. Additionally, an analytical device is needed to control and monitor data from the sensor. A potentiostat fulfills this role by conducting electrochemical analyses. When combined with an electrochemical sensor, the potentiostat creates a fully integrated electrochemical sensing device, sampling and sending data to the user[19]. Both components must work together to ensure accurate measurement results for detecting the target biomarker or analyte. While potentiostat Integrated Circuits (ICs) are available, they require significant power (tens of milliwatts) and rely on a microcontroller for data processing and transmission. An example is the AD5940 IC from Analog Devices, which contains a full potentiostat circuit but requires a separate Microcontroller Unit (MCU) for data handling. While suitable for desktop devices without size or power restrictions, ICs like these are less ideal for wearable devices that need to be compact and energy efficient. Furthermore, wearable analytical devices must match the precision of benchtop instruments while maintaining a small form factor, no larger than a wristwatch. Power consumption also needs to be minimized so the device can run on a battery for several days without recharging. Given the limited size of wearable devices, battery capacity is constrained, with lithium-ion (Li-ion) batteries typically offering only 10-20 mAh, depending on the available space.

This work presents an electrochemical analytical device (i.e., a potentiostat) designed using STMicroelectronics' low-power STM32L476 microcontroller for digitizing analog signals and signal processing. The potentiostat is capable of running Electrochemical Impedance Spectroscopy (EIS) and Cyclic Voltammetry (CV) techniques. Wireless communication is enabled through Silicon Labs' EFR32BG13 Bluetooth System on a Chip (SoC) with an integrated Bluetooth transceiver. The STM32L476 microcontroller, based on the ARM M4 core, includes a built-in analog-to-digital converter (ADC) and a digital-to-analog converter (DAC). It features three independent ADCs with a sampling rate of up to 5 mega samples per second (Msps) and hardware oversampling capabilities. The DAC offers 12-bit resolution with a low-power sample-and-hold circuit. The STM32L476 also supports low-power modes, with Stop 1 and Stop 2 modes providing full data retention and fast wake-up times of 5μs and 4μs, respectively. The design integrates an instrumentation amplifier as an electrometer, a transimpedance amplifier (TIA), a control amplifier, and an input multiplexer to manage multiple working electrodes. The device can select the appropriate working electrode, supply optimal current to the sensor, periodically sample sensor currents and voltages, and calculate impedance. Basic signal processing, including averaging and complex impedance calculations, is carried out by the STM32L476 MCU. For performance evaluation and testing, serial communication via Universal Serial Bus (USB) is used instead of wireless communication. The collected data and computed values are transmitted to a host computer through the UART interface for testing. The performance of the developed potentiostat was assessed in comparison to an existing commercial potentiostat device, EmStat from BASI Inc. Furthermore, the prototype includes hardware support for Bluetooth Low Energy (BLE) communication, which will be added in the future to enable wireless data transmission to cloud services for more advanced analysis.

## Literature Review

**Existing wearable potentiostats:** As electronic components become more advanced and compact, the development of wearable sensing devices becomes increasingly feasible. Modern integrated circuits, such as microcontrollers, analog-to-digital converters, and digital-to-analog converters, are now smaller and more energy-efficient than similar devices from 10 to 15 years ago. Most potentiostats developed in the last five years have been optimized for cyclic voltammetry (CV) measurements and steady-state voltage for amperometric analysis. CV is used to identify anodic and cathodic peak currents, with the voltage at the peak current serving for amperometric measurements[20]. There are various approaches to designing the analog front-end of these devices. Some potentiostats use integrated analog front-end solutions, while others rely on operational amplifiers. Integrated designs are more compact but offer less flexibility compared to discrete designs. Recently, some research groups have developed advanced potentiostats utilizing pulse voltammetry and impedance spectroscopy techniques. Despite advancements, wearable sensing instrumentation faces challenges related to size and power consumption.

One example is Ahmad et al.'s open-source wearable potentiostat, "KAUSTat"[21]. This low-power, wireless device allows users to set parameters such as start voltage, final voltage, step voltage, and the number of measurement cycles. The design includes analog potentiostat circuits (TIA, control amplifier, and electrometer), a digital-to-analog converter (DAC), an analog-to-digital converter (ADC), and Bluetooth connectivity. Powered by a battery, it uses voltage regulators to maintain appropriate power levels and transmits data to a smartphone via Bluetooth Low Energy (BLE). A smartphone app controls the potentiostat and sets its parameters. Comparative testing showed that KAUSTat's measurements were similar to those of a commercial potentiostat.

Another smartphone-based potentiostat is built around the ESP32 Espressif 32 chipset[22], using a modified gas sensor front-end (LMP91000EVM) from Texas Instruments. Evaluated using potassium ferricyanide solution with cyclic voltammetry (CV), this system showed a good electrochemical response compared to a commercial workstation. Janyasupab et al. noted that this cost-effective design can be prototyped for around USD 80 and mass-produced at a lower cost.

A flexible wireless system developed through a collaboration between North Carolina State University and the University of North Carolina at Chapel Hill is designed to analyze biomarkers in sweat [23]. This flexible potentiostat is shaped like a wristwatch and powered by a 3.7 V lithium polymer battery. The potentiostat is controlled by a Bluetooth Low Energy (BLE) system-on-chip (CC2642, from Texas Instruments), which connects to an electrochemical analog front end (AFE) (AD5941, Analog Devices) that includes the bias and transimpedance amplifier circuits necessary for amperometric sensing. The SoC can control the sampling rate, bias voltage, and gain of the electrochemical cell. Once initialized, the AFE reads data from the electrochemical cell, stores the amperometric current data in a buffer, and then the SoC retrieves this data and transmits it via BLE.

Al-Hamry et al. developed a low-cost, portable impedance analyzer using an STM32 microcontroller[24]. Their system employs two DACs and two differential ADC channels for simultaneous current and voltage sampling. For impedance spectroscopy, one DAC generates a DC offset, while the other produces a sine wave. Data is transferred via a serial link to a host PC, and the system analyzes current and voltage samples using Discrete Fourier Transform (DFT) to obtain impedance magnitude and phase. The prototype can measure impedances below 10kΩ and capacitances below 10μF, with a frequency range from 0.1Hz to 100kHz. Accuracy depends on the resistance-to-capacitance ratio.

Another group led by Huang from the National University of Tainan, Taiwan, developed a portable potentiostat using two 8051-based 8-bit microprocessors[25]. This device features 12-bit DACs to generate excitation waveforms and DC offsets, and a 12-bit ADC to digitize current and voltage signals. Designed to perform CV scans, it does not support impedance spectroscopy. The CV scan results closely matched those of a commercial potentiostat.

**Techniques for reducing measurement errors:** In addition to various wearable potentiostat designs capable of performing cyclic voltammetry (CV) and constant voltage bias, some research groups are exploring new techniques and algorithms to enhance measurement accuracy and reduce error rates. A potentiostat developed by the Department of Electrical and Computer Engineering at Iowa State University shows promising results. This potentiostat uses operational amplifiers for signal conditioning, potentiostat functionality, output signal conditioning, and single-ended to differential conversion for the ADC interface. Data acquisition, signal processing, and transfer are managed by the C2000 microcontroller from Texas Instruments. It employs a Multiset Differential Pulse Voltammetry (MS-DPV) technique, which improves electrochemical sensing precision by sampling currents near the rising and falling edges of each pulse. These samples are processed using algorithms to generate differential measurements, allowing multiple readings with the same resources typically used for one. This process reduces error by a factor of $\sqrt{N}$, where N is the number of measurements, yielding a 12% error reduction for a four-set DPV[26].

Impedance spectroscopy, however, remains one of the most challenging measurement methods for wearable potentiostat systems. This technique requires precise analog circuits and microcontrollers, with high-speed sampling and complex computations adding difficulty in miniaturized instruments. Multiple research groups have attempted solutions for accurate impedance and phase measurements. Yu et al.'s group measured phase using an Exclusive OR (XOR) gate followed by a low-pass filter[27], with the resulting DC voltage representing a phase shift from 0° to 180°. Impedance magnitude was measured with an open-loop peak detector, providing accurate results, though phase measurement data was not discussed. The group tested impedance values ranging from 970Ω to 1620Ω, with an error rate of up to 2.71%. Another potentiostat, based on Analog Devices' AD5933 impedance analyzer IC, operates within a frequency range of 10 Hz to 10 kHz[28]. Though wider than Yu et al.'s design[27], this range is still insufficient for low-frequency processes like solution diffusion at electrode surfaces (below 10 Hz).

Pruna et al. designed a low-cost potentiostat ($300 per unit) capable of impedance spectroscopy, using a PIC32MX795F512L microcontroller for communication and signal processing, and a PIC24FJ128GC010 for analog-to-digital conversion and waveform generation[29]. Its frequency range is 0.1 Hz to 10 kHz, and its results were comparable to commercial potentiostats, though measurements at 10 kHz were highly noise-sensitive. The miniature low-power potentiostat prototype developed in this thesis offers a broader bandwidth for impedance spectroscopy than most of the above-mentioned designs, except for the AD5933, which has a programmable excitation voltage up to 100 kHz but a narrower impedance measurement range (1 kΩ to 1 MΩ). This prototype can provide excitation voltage in the range of 100 Hz to 50 kHz for Electrochemical Impedance Spectroscopy (EIS), with the potential to expand from 0.1 Hz to 100 kHz by modifying the firmware and adjusting the low-pass filter cutoffs. There is no high-pass filter in the design, so a firmware update is required to change the frequency range. All parameters, including start and end frequencies, frequency increments, and the number of measurements for averaging, are software-controlled. Averaging helps reduce noise-induced errors, particularly when sensor impedance is high and current is low.

In 2020, Gücin et al. published research on impedance spectrum measurement using cross-correlation[30], focusing on evaluating battery performance. Their system conducts impedance spectroscopy without disconnecting the battery, using charging current and battery voltage for complex measurements. White noise serves as the excitation signal, and the system uses autocorrelation and cross-correlation of input and output signals to perform a Fourier transform, obtaining the battery's frequency response, which is used to derive its complex impedance. The following mathematical equation summarizes the process of obtaining the frequency response.

$$H(j\omega) = \frac{R_{uy}(j\omega)}{R_{uu}(j\omega)}$$

Where $R_{uy}(j\omega)$ is Fourier transform of cross-correlation sequence of input and output signals while $R_{uu}(j\omega)$ is a Fourier transform of the autocorrelation of the input signal. This system requires fast computation capabilities and may not be a good fit for low-power applications. The experimental setup is realized using a field programmable gate array (FPGA) integrated circuit.

This impedance measurement technique could be considered for chemical sensor analysis if it can be adapted for use in low-power devices. Yang et al.'s group proposed an integrated solution for impedance spectroscopy, incorporating an array of working electrodes, potentiostat analog circuits, an ADC, phase and peak detectors, memory, and a communication interface[31]. They explored two approaches for accurate impedance measurements. The first method involves using pseudo white noise and Fast Fourier Transform (FFT) to determine the sensor's frequency response. The second approach, based on a Frequency Response Analyzer (FRA), generates results for one frequency at a time. By mixing the sinusoidal signal from the potentiostat output with sine and cosine signals, the frequency response is shifted to DC to obtain magnitude and phase data. The FRA-based method requires less computational power than FFT, making it a more suitable choice for miniature, integrated impedance spectroscopy systems.

# Hardware Design

The designed potentiostat is composed of three main subsystems: analog potentiostat circuits, a microcontroller for analog-to-digital conversion, and a wireless communication module (BLE). Figure 1 provides a simplified diagram showing how these subsystems are connected.

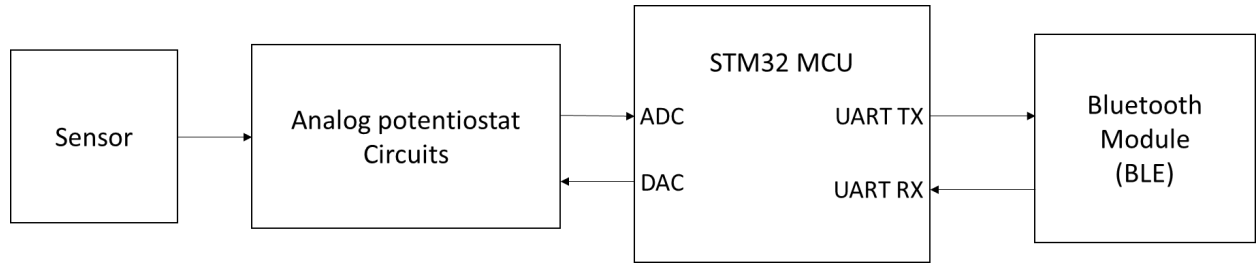

Figure 1: Potentiostat block diagram

The analog potentiostat circuit is made up of three sub-circuits, as illustrated in Figure 2: a trans-impedance amplifier (TIA), an electrometer, and a control amplifier. This circuit operates in a closed-loop configuration with the control amplifier, which is connected to the counter electrode of a three electrode-based biochemical sensor. The excitation voltage is supplied via the DAC, and the electrometer's input terminals are attached to the sensor's reference and working (sense) electrodes. The electrometer's output, representing the voltage difference between the reference and working electrodes (i.e., the sensor's voltage drop), is fed back to the control amplifier and also sent to the ADC for digitization. The TIA converts the current flowing through the sensor into a voltage, and its output is connected to another ADC input. Figure 2 shows a simplified representation of the potentiostat's circuit.

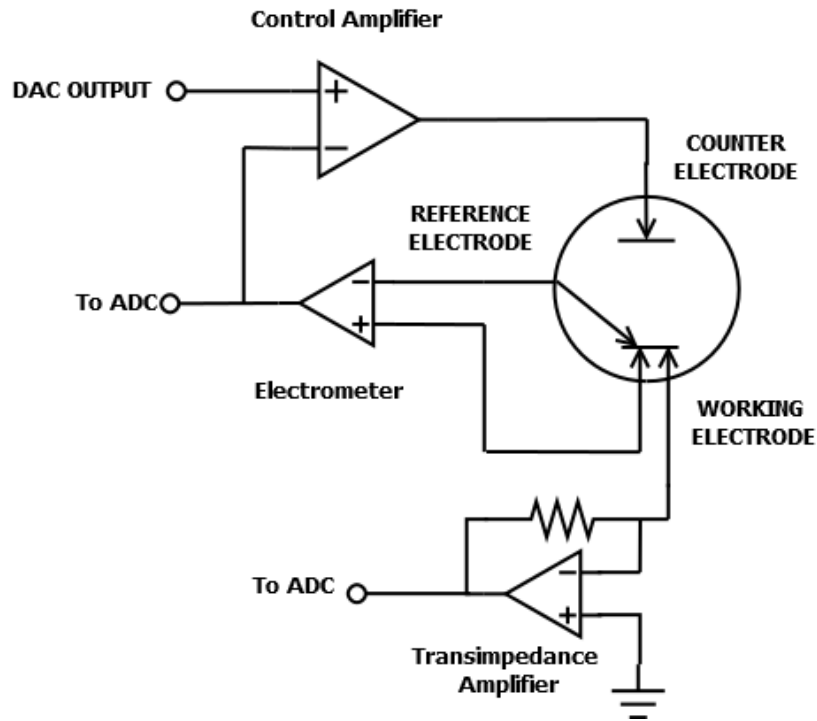

Figure 2: Analog potentiostat subcircuits

This low-power potentiostat is specifically designed for integration with wearable sensors, and several constraints must be met to achieve this. One key factor is its low-power design, which allows it to be battery-operated. The device operates on a supply voltage between 3.5V and 4V, aligning with the nominal 3.7V of a lithium-ion battery. Additionally, it does not require negative supply voltages, which reduces the need for extra voltage regulators and simplifies the overall design. Compared to other solutions, this implementation is relatively straightforward and can be realized using a standard ARM core-based microcontroller, making it suitable for battery-powered operation. The design also supports low-power functionality and offers flexibility for adding new features through firmware updates.

## Digital Signal Processing

**Impedance spectroscopy:** To run EIS, we developed the following process flow. Analog-to-digital conversion is carried out using the microcontroller's integrated 12-bit ADCs, requiring two channels to sample outputs from the electrometer and transimpedance amplifiers. The microcontroller's built-in 12-bit DAC generates an excitation signal for the potentiostat circuit. Depending on the measurement modality, the microcontroller processes the signals to derive voltage and current readings. For impedance spectroscopy, the excitation signal is a sinusoidal waveform, with voltage and current represented as root mean square values. Impedance is calculated from these measurements, and phase is determined using cross-correlation between voltage and current values, with a selected phase resolution of one degree. To achieve this resolution, the outputs must be oversampled to obtain 360 samples per excitation cycle. This necessitates adjustments to the ADC sampling rate during impedance measurements to minimize memory usage at low excitation frequencies. Accurate phase shift calculations depend on using the correct sampling rates, which the firmware continuously monitors and adjusts to maintain one-degree resolution across the measurement range. However, this one-degree resolution is only attainable at excitation frequencies below 11.4 kHz; above 20 kHz, the resolution is approximately two degrees. A simplified software flow chart for measuring the magnitude and phase of impedance is shown in Figure 3.

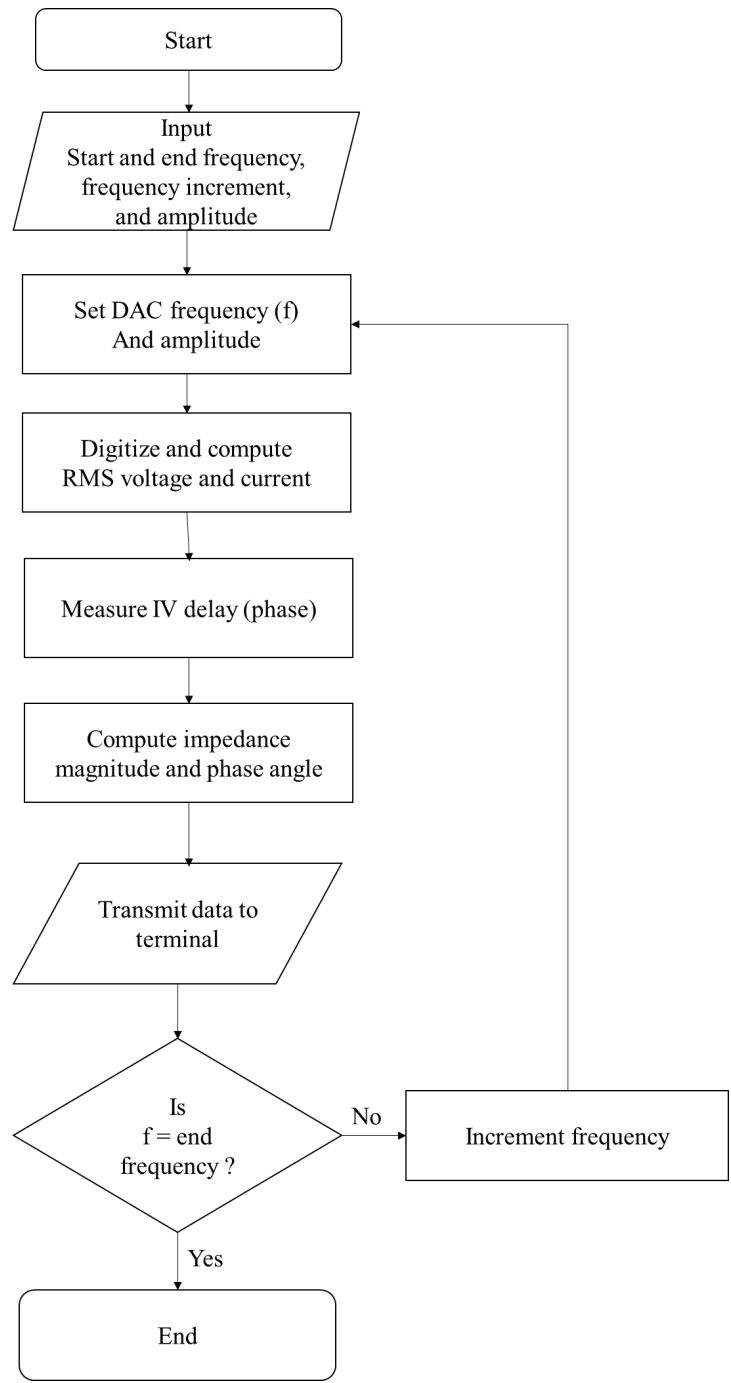

Figure 3: Firmware flow diagram (EIS scan)

The algorithm for converting ADC samples to RMS measurements is widely recognized in digital signal processing. Before obtaining the root mean square values, it is essential to remove the DC offset from the acquired samples, which is calculated as the mean of the data samples. The current and voltage samples are converted to RMS values by squaring each sample and summing these squared values in an accumulator. The RMS value is then calculated by taking the square root of the accumulated sum and dividing it by the number of samples[32]. This method is applied to both voltage and current in impedance spectroscopy. This algorithm can be described mathematically using a single equation.

$$V_{rms} = \sqrt{\frac{1}{N}\sum_{n=0}^{N-1} x_n^2} \qquad (1)$$

Where N is the number of samples and $x_n$ is the instantaneous voltage measurement obtained by the ADC. Raw ADC count is converted to the instantaneous voltage using the arithmetic operation shown in equation 2.

$$V = \frac{V_{ref}}{4095} \times ADC\ count \qquad (2)$$

Where $V_{ref}$ is an ADC reference voltage (3.3V) and 4095 is ADC full-scale count (12-bit ADC). Since the potentiostat employs a transimpedance amplifier to convert current into voltage, the current can also be expressed as a voltage, using the same algorithm to derive the current from the sensor. To convert voltage back to current, the RMS voltage output from the TIA is divided by the feedback resistance in the transimpedance amplifier circuit. Due to the inverting nature of the TIA, the resulting current is phase-shifted by 180 degrees relative to the voltage.

The complex impedance value is determined by measuring the phase shift between the voltage and current. The voltage and current samples from the ADC, used for RMS calculations, also contribute to phase measurement. The outputs from the transimpedance amplifier and the electrometer are oversampled beyond the Nyquist rate. For impedance spectroscopy, the sample rate must be dynamically adjusted to accommodate the changing frequency of the excitation signal, with phase resolution depending on the number of samples taken within one cycle. The goal is to obtain approximately 360 samples per cycle, yielding a 1° phase resolution. The phase shift is calculated through cross-correlation between the voltage and current. This cross-correlation is executed across the entire phase shift range from 0° to 360°. The lag corresponding to the maximum cross-correlation value indicates the delay between the current and voltage, measured in samples. This lag is then converted into a phase shift by multiplying it with the degrees per cycle to express the lags in degrees. The cross-correlation algorithm can be described using the following equation:

$$R_{x,y}[k] = \sum_{l=-\infty}^{\infty} x[l]y[k+l] \tag{3}$$

Where x[l] is the electrometer voltage sample and y[l] is the TIA sample[33]. Conversion from lags to phase shift is performed using the following equation.

$$\Phi = k\left(\frac{360}{N}\right) - 180° \tag{4}$$

Where Φ is phase shift, k is lag, N is the number of samples per cycle, and 180° is subtracted due to the inverting nature of TIA.

**Cyclic voltammetry:** Cyclic voltammetry (CV) scanning is less complex than impedance scanning. To run CV, we developed the following process flow. The input parameters of CV scan include the voltage change rate in volts per second (V/s), the starting voltage, and the ending voltage. The starting voltage serves as the initial DAC voltage, while the ending voltage represents the final DAC value. The DAC updates according to the specified voltage change rate. Once the DAC is activated, the ADC begins sampling the outputs from the electrometer and transimpedance amplifier (TIA). These samples are stored in a transmit buffer using the direct memory access (DMA) feature of the microcontroller, which helps prevent potential data loss during acquisition. Data collection continues until all current and voltage measurements are obtained within the specified voltage range. Once the scan is complete, the stored samples are converted into voltage and current values. Following each conversion, the current and voltage data are transmitted to the user through a serial communication interface. Figure 4 illustrates the firmware flow diagram for the CV scan.

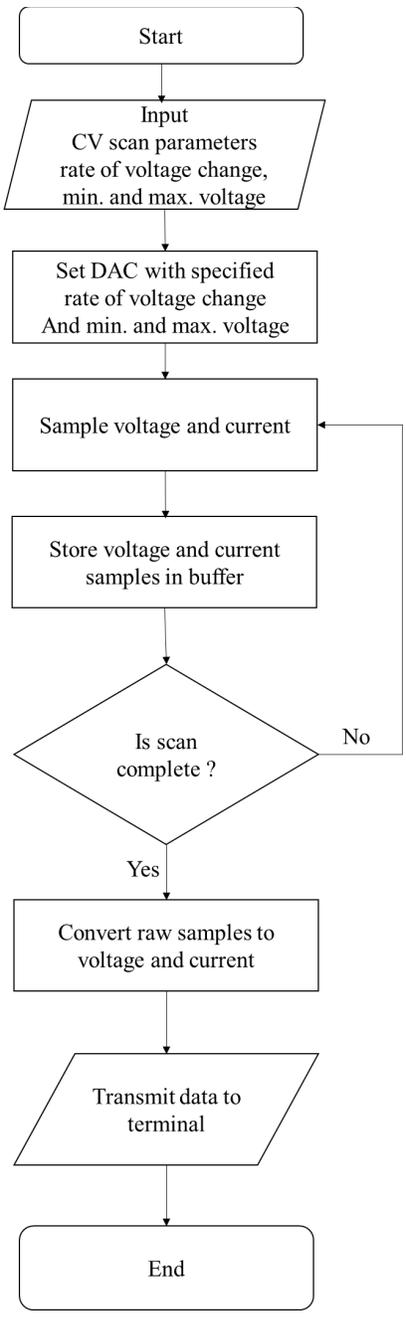

Figure 4: Firmware flow diagram (CV scan)

## Materials and Methods

**Potentiostat circuit board:** In this research, a prototype potentiostat board was designed and assembled. The board includes an analog potentiostat circuit and a Bluetooth communication subcircuit. A separate board houses the microcontroller, with both boards connected via wires and shielded cables for the noise-sensitive analog signals. The prototype potentiostat is powered by a benchtop power supply. Figure 5 below displays a three dimensional (3D)-rendered model of the potentiostat board.

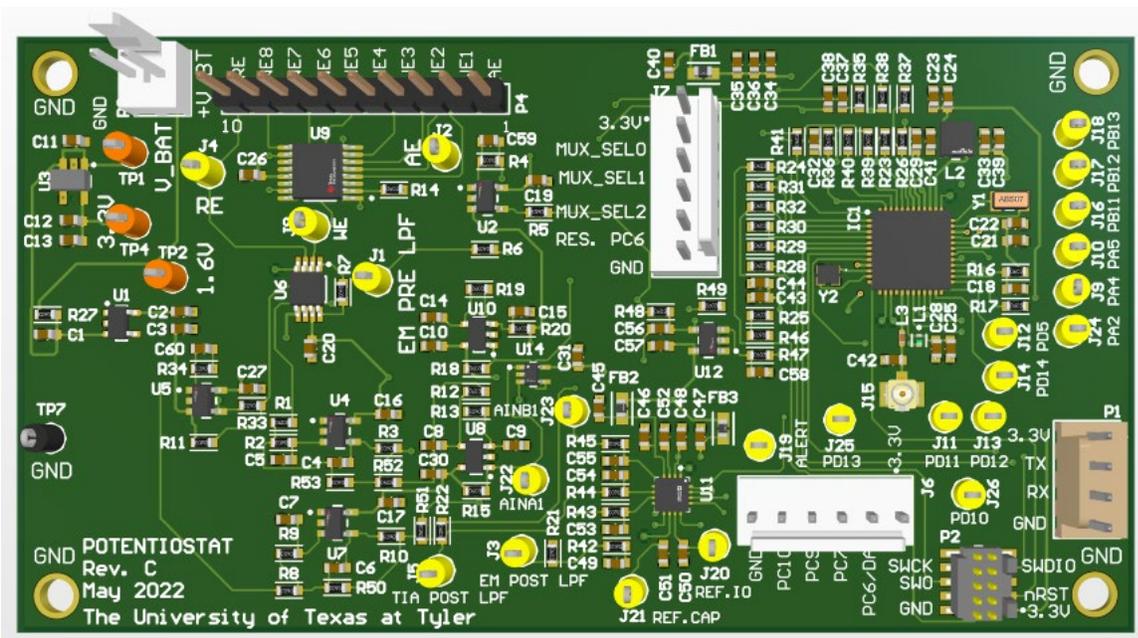

Figure 5: Potentiostat PCB 3D render

The potentiostat board includes voltage regulators for the analog circuits, potentiostat subcircuits with an input multiplexer, and a Bluetooth subcircuit. The microcontroller board is powered through a USB connection. Additionally, there is a third adapter board for communication, which acts as a UART-to-USB bridge using the FTDI FT232RL chip. This adapter board is only used during testing and will be unnecessary once Bluetooth communication is fully functional. Figure 6 shows the experimental setup, featuring the assembled potentiostat board (left) and the STM32L476 evaluation board (right) during testing. The microcontroller board is an evaluation kit for the STM32L476 microcontroller. Shielded short cables are used to connect the transimpedance amplifier (TIA) and electrometer to the ADC inputs of the STM32L476.

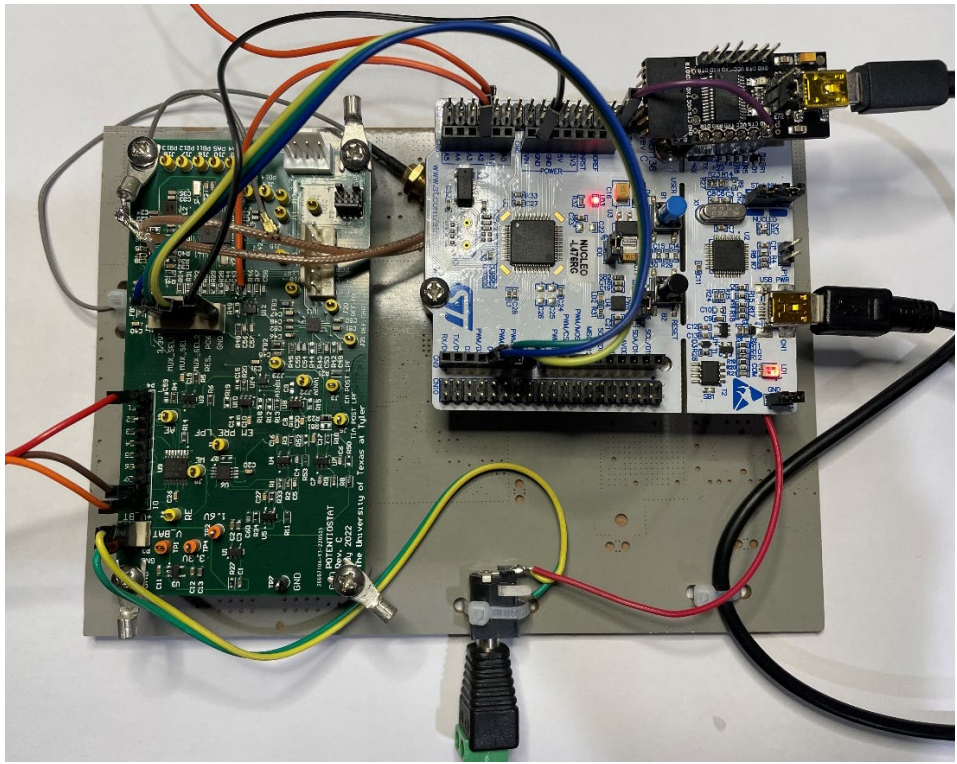

Figure 6: Assembled potentiostat board

**Custom PC program for communication:** The commands needed to control and configure the potentiostat are 22 bytes long, making testing and evaluation challenging without a custom program to send these commands. To simplify this process, a PC test application was developed to send commands and collect data. The application enables users to set parameters such as scan frequency range, frequency increment, excitation signal amplitude, CV rate, start and end CV scan voltages, and other evaluation commands. Additionally, the application allows potentiostat data to be saved on a computer in a tab-delimited format for further analysis, which can be visualized and processed using Excel and MATLAB. Figure 7 displays the user interface of the developed PC application

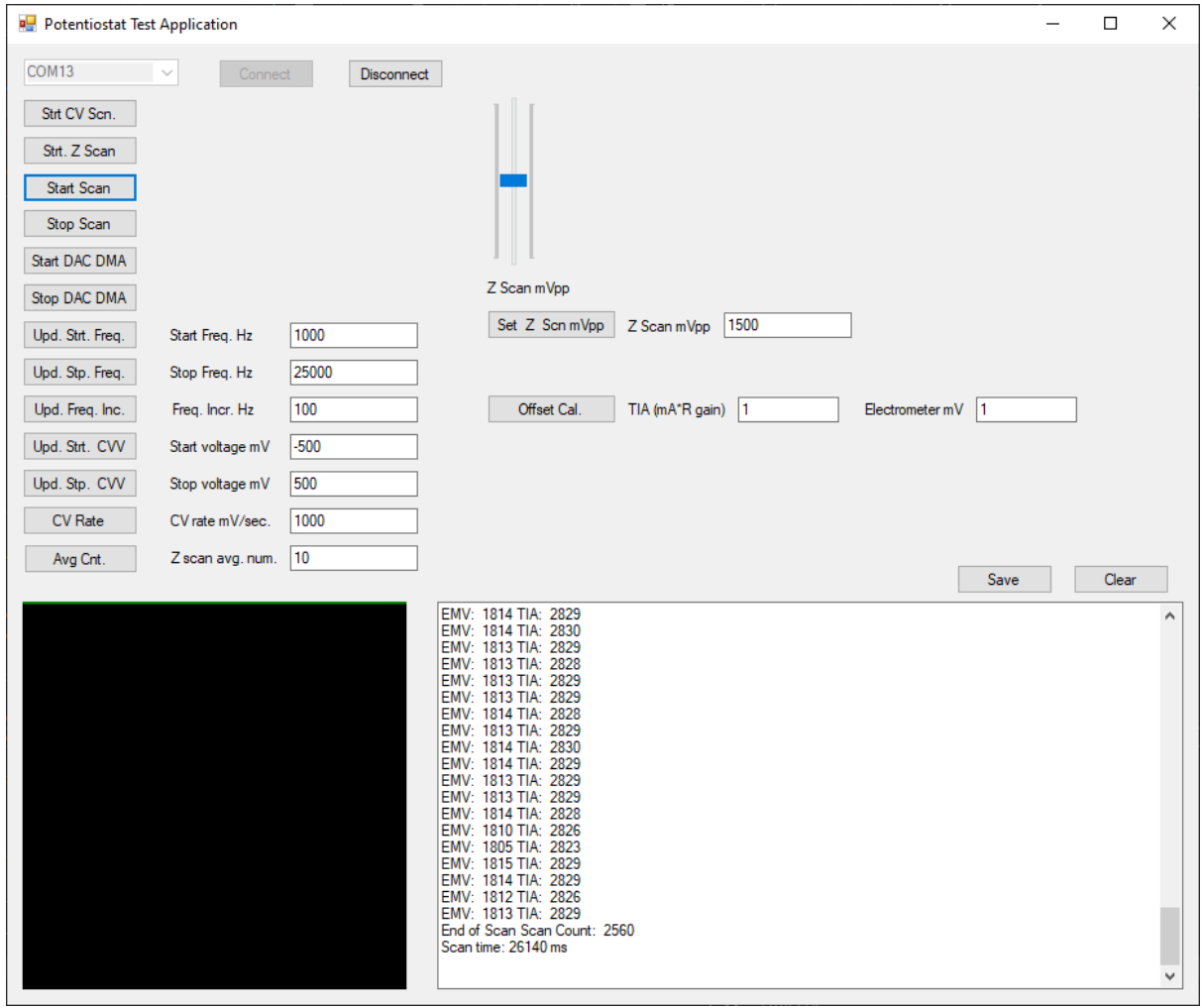

Figure 7: Screenshot of the test PC application

**Experimental data analysis:** Tests were conducted to assess the performance of the potentiostat prototype, focusing on power consumption, measurement accuracy, and sensitivity, as the aim is to make the device wearable. Validation involved a series of tests in the electronics laboratory.

Key factors for wearable suitability include power consumption, size, and measurement accuracy. Currently, the prototype is not compact enough for wearable use. This is the first functional version, and the research primarily aims to enhance its performance, particularly in terms of accuracy and power efficiency. The accuracy of cyclic voltammetry (CV) and impedance measurements was verified using resistive and capacitive circuits with known impedances. The test results were compared with expected values, allowing precise error determination and establishing an accuracy baseline. Additionally, the potentiostat was calibrated for phase offset due to ADC front-end multiplexer delays and voltage offsets. The percent error in impedance measurements is given below.

$$Percent\ Error = \frac{Measured\ Impedance - Expected\ Impedance}{Expected\ Impedance} 100\%\ [\%] \qquad (5)$$

The phase error is measured as a difference between the measured value and the expected value.

$$Phase\ Error = Measured\ Phase - Expected\ Phase\ [degrees] \qquad (6)$$

The expected impedance magnitude is calculated using the equation shown below.

$$|Z| = \frac{1}{\sqrt{\frac{1}{R^2} + (2\pi f C)^2}} \qquad (7)$$

Where f is excitation frequency, R is resistor value in ohms, and C is capacitor value in Farads. The phase shift is calculated using the following equation.

$$\emptyset = \tan^{-1}(-2\pi f RC) \qquad (8)$$

The tests mentioned above were conducted under controlled conditions using laboratory equipment to simulate biochemical cells. The evaluation tools included capacitor and resistance decade boxes, oscilloscopes, and a digital multimeter. An LCR meter was employed to measure capacitance.

**Impedance spectroscopy tests:** An impedance spectroscopy test was carried out by connecting a biochemical sensor to the developed potentiostat. The same test was then repeated under identical conditions using a commercial potentiostat (EmStat, BASI Inc.), and the data was saved for comparison. The test results from the developed potentiostat were compared with those from the commercial device. A MATLAB script was used to plot the data from both potentiostats for visualization. The detailed validation process, along with the test data, is discussed in the Results and Discussion section.

**Cyclic voltammetry tests:** A cyclic voltammetry (CV) test was performed to compare the data from the developed potentiostat with that from a commercial potentiostat. The test was conducted using the same sensor and analyte concentration for both devices. The CV scan parameters, including the rate, start voltage, and voltage range, were identical for both potentiostats. The collected data was then compared by plotting the results from both devices on a single graph using MATLAB for visualization.

# Results and Discussion

**Impedance accuracy and range test:** This experiment establishes a baseline for measurement accuracy, with baseline data collected using pure resistors. Table 1 presents the TIA current and electrometer voltage outputs from the developed potentiostat, measured with an oscilloscope. The recorded voltage and current values were compared to those obtained from the commercial potentiostat.

Table 1. Resistance measurement data at fixed frequency 9920 Hz

| Actual Resistance | Measured with Developed Potentiostat | | | Measured with Commercial Potentiostat | | |
|---|---|---|---|---|---|---|
| | $I_{rms}$ [mA] | $V_{rms}$ [mV] | Resistance [Ω] | $I_{rms}$ [mA] | $V_{rms}$ mV | Resistance [Ω] |
| 1000 | 0.068 | 63.6 | 935.29 | 0.062 | 62.66 | 1003 |
| 5000 | 0.029 | 133 | 4586.21 | 0.0266 | 131.867 | 4956 |
| 10000 | 0.0185 | 173 | 9351.35 | 0.0175 | 171.379 | 9793 |
| 50000 | 0.0065 | 309 | 47538.46 | 0.0065 | 299.66 | 45898 |
| 100000 | 0.005 | 470 | 94000.00 | 0.0051 | 460.796 | 89966 |
| 150000 | 0.0033 | 472 | 143030.30 | 0.0036 | 461 | 126247 |
| 200000 | 0.0025 | 472 | 188800.00 | 0.003 | 461 | 154659 |

Table 2 displays the calculated percent errors from the measurements presented in Table 1. These percent errors are derived using Equation 5 from the methods section. Two types of percent error calculations are included. The first, labeled as analog percent error, indicates the error in the impedance magnitude by comparing the voltage and current measured at the output of the developed potentiostat with the actual resistance value (as indicated on the resistance decade box). The second, referred to as digital percent error, is calculated by comparing the final result reported by the software with the actual resistance value.

Table 2. Analog circuit and digital circuit percent error

| Actual | Measured Resistance [Ω] (Analog Potentiostat Circuit) | Percent error (Analog) | Resistance Reported in Software [Ω] | Percent Error (Digital) |
|---|---|---|---|---|
| 1000 | 935 | -6.47 | 1003 | 0.30 |
| 5000 | 4586 | -8.28 | 4956 | -0.88 |
| 10000 | 9351 | -6.49 | 9793 | -2.07 |
| 50000 | 47538 | -4.92 | 45898 | -8.20 |
| 100000 | 94000 | -6.00 | 89966 | -10.03 |
| 150000 | 143030 | -4.65 | 126247 | -15.84 |
| 200000 | 188800 | -5.60 | 154659 | -22.67 |

The analog circuit shows a percent error between -4.65% and -8.3%, with no clear trend between the error and the applied resistance value. Table 2 indicates that the resistance measured by the analog circuit is consistently lower than the actual resistance, likely due to a gain error in the circuit. The software-reported resistance, however, shows a pattern where the percent error increases with higher resistance values. Despite the potentiostat being calibrated for offset before measurements, the error trend suggests that an ADC gain error is the likely cause.

The accuracy of complex impedance measurements is validated using a capacitor decade box by testing various capacitor values in parallel with 1kΩ and 10kΩ resistors. The results are compared with theoretical impedance magnitude and phase shift values, calculated using equations 7 and 8 from the methods section. Measurements are conducted at a fixed test frequency of 10,080 Hz. Table 3 shows a phase error of -2.53 degrees and a magnitude error of approximately 6% for a 10kΩ resistor and 8.2nF capacitor in parallel. When testing 1kΩ resistance with capacitors ranging from 4.3nF to 10nF, the impedance magnitude error is between ~1% and 4%, while the phase error is slightly larger by 1 to 2 degrees compared to the 10kΩ results. For 1kΩ resistance, the phase error ranges from ~1 to 2.8 degrees.

Table 3: RC circuit impedance scan data

| R[Ω] | C [F] | C[nF] | Xc [Ω] | Calculated \|Z\| [Ω] | Calculated Phase Shift φ [deg.] | Measured \|Z\| [Ω] | Measured Phase Shift φ [deg.] | Phase Error [deg.] | Impedance Magnitude % Error |
|---|---|---|---|---|---|---|---|---|---|
| 1000 | 1.00E-08 | 10.000 | -1578.9 | 844.81 | -32.35 | 870 | -30.51 | -1.84 | 2.98 |
| 1000 | 8.20E-09 | 8.200 | -1925.5 | 887.46 | -27.44 | 923 | -25.3 | -2.14 | 4.01 |
| 1000 | 6.74E-09 | 6.740 | -2342.6 | 919.71 | -23.12 | 951 | -20.27 | -2.85 | 3.40 |
| 1000 | 5.55E-09 | 5.550 | -2844.9 | 943.41 | -19.37 | 953 | -20.27 | 0.90 | 1.02 |
| 1000 | 4.30E-09 | 4.300 | -3671.9 | 964.86 | -15.23 | 973 | -16.86 | 1.63 | 0.84 |
| 10000 | 8.20E-09 | 8.200 | -1925.5 | 1890.78 | -79.10 | 2000 | -76.57 | -2.53 | 5.78 |
| 10000 | 6.74E-09 | 6.740 | -2342.6 | 2280.86 | -76.82 | 2400 | -76.57 | -0.25 | 5.22 |
| 10000 | 5.55E-09 | 5.550 | -2844.9 | 2736.32 | -74.12 | 2900 | -73.16 | -0.96 | 5.98 |
| 10000 | 4.40E-09 | 4.400 | -3588.5 | 3377.57 | -70.26 | 3700 | -69.75 | -0.51 | 9.55 |

**RC circuit impedance spectroscopy test:** Accuracy is assessed using a simulated sensor created with resistors and capacitors arranged in both series and parallel configurations. The potentiostat's measurement accuracy was confirmed by comparing the measured values to the expected calculated values.

In this test, the circuit consists of a parallel RC combination (a 10kΩ resistor and a 33nF capacitor) in series with a 560Ω resistor. The RC circuit diagram is displayed in Figure 8 below, with terminal connections clearly labeled. The terminal marked CE corresponds to the control amplifier connection, WE is the working electrode connection, and RE is the reference electrode connection.

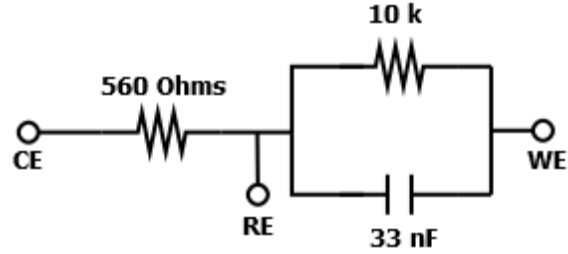

Figure 8: EIS scan RC test circuit connection diagram

The developed potentiostat was compared to a commercial benchtop potentiostat, with both devices configured to perform impedance spectroscopy measurements (impedance scans) from 100Hz to 25kHz. The low-power potentiostat we developed conducted the impedance scan in increments of 50Hz, while the commercial potentiostat scanned at fewer frequency points. To facilitate comparison, a MATLAB script was developed to select data points from the low-power potentiostat that closely matched the excitation frequencies of the commercial device. The collected magnitude and phase data were plotted on the same graph in Figure 9 for easier comparison. The graph shows that the impedance magnitude curve from the low-power potentiostat nearly overlaps with that of the commercial potentiostat across the frequency range of 100Hz to 25kHz. At lower excitation frequencies (100Hz to 2000Hz), the impedance magnitude error is larger compared to frequencies above 2000Hz. The phase measurements exhibit a similar trend throughout the scan, with a phase error of $\pm 5°$ for most data points. However, there are phase errors exceeding 10 degrees in the ranges of 1.5kHz to 1.6kHz and 150Hz to 250Hz.

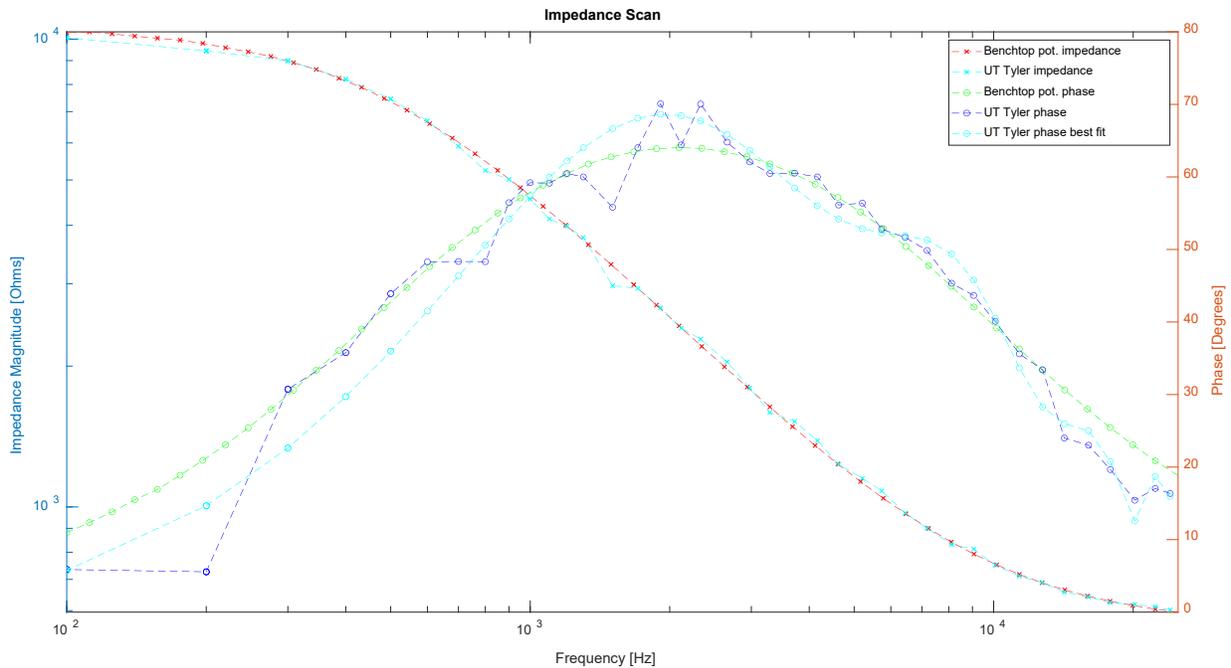

Figure 9: RC EIS scan, showing a comparison between benchtop and developed (UT Tyler) potentiostats

**Redox circuit impedance spectroscopy test:** In this test, the developed potentiostat was evaluated using the RedOx circuit supplied with the commercial benchtop potentiostat, which simulates a RedOx chemical cell. The low-power potentiostat performed a frequency scan in 50Hz increments, ranging from 100Hz to 25kHz. The data collected were compared with those from the benchtop potentiostat, and a MATLAB script was used to select data points from the low-power potentiostat that corresponded to the excitation frequencies of the commercial device. The magnitude and phase data were plotted on the same graph in Figure 10 for comparison. The results indicate that the impedance magnitude shows larger errors at the beginning of the scan (100Hz to 500Hz) and at the higher end (above 10kHz) when compared to the commercial potentiostat. Additionally, the phase measurement exhibits significant errors in the frequency ranges of 1.5kHz to 1.6kHz and 6kHz to 9kHz.

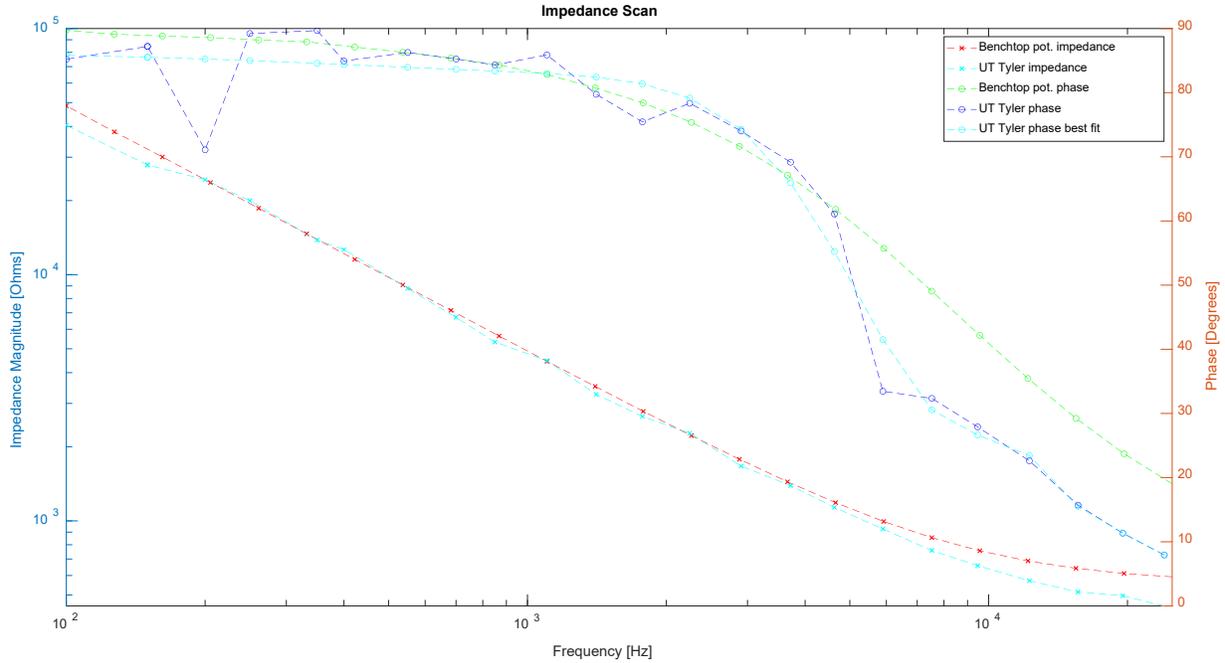

Figure 10: RedOx circuit EIS scan, showing a comparison between benchtop and developed (UT Tyler) potentiostats

**Laser-induced graphene (LIG) electrode impedance spectroscopy test:** This test compares measurements taken with a commercial potentiostat to those obtained using our potentiostat when interfaced with a laser-induced graphene (LIG) electrode developed in our prior work[34]. Figure 11 shows that at a low frequency of 100 Hz, the impedance magnitude measured with the developed potentiostat is 23kΩ, while the commercial benchtop potentiostat measures 19.6kΩ, resulting in an error of approximately 15.6%. The highest error, 24%, occurs at 500 Hz. Additionally, the impedance scan in Figure 11 reveals that the error in impedance magnitude is frequency-dependent. At 10 kHz, the error is -9.8%, indicating that the measured impedance is lower than expected. This trend suggests a relationship between error and excitation frequency, as seen in Figure 12. Furthermore, the phase values measured with the developed potentiostat differ from expected values by ±4 degrees, with two exceptions: at 200 Hz, the phase differs by

12 degrees, and at 1.5 kHz, the phase is 11 degrees off. These phase measurement discrepancies at 200 Hz and 1.5 kHz are consistent across all tests using the sensors.

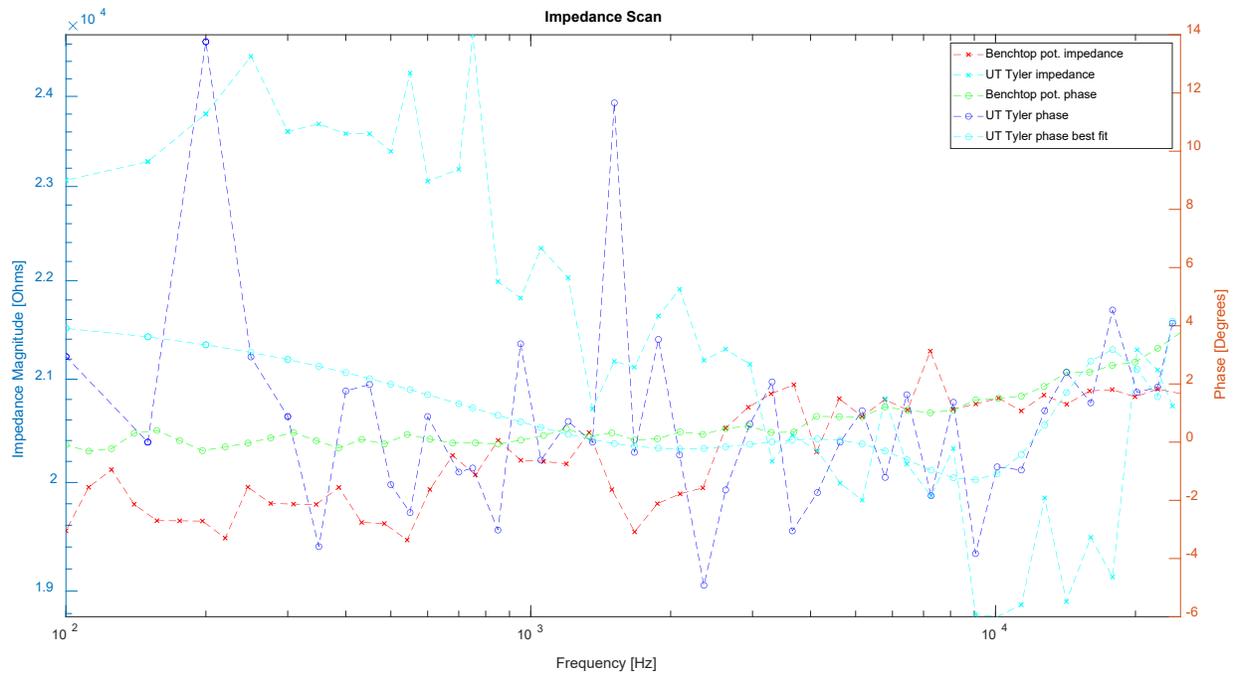

Figure 11: LIG electrode EIS scan, showing a comparison between benchtop and developed (UT Tyler) potentiostats

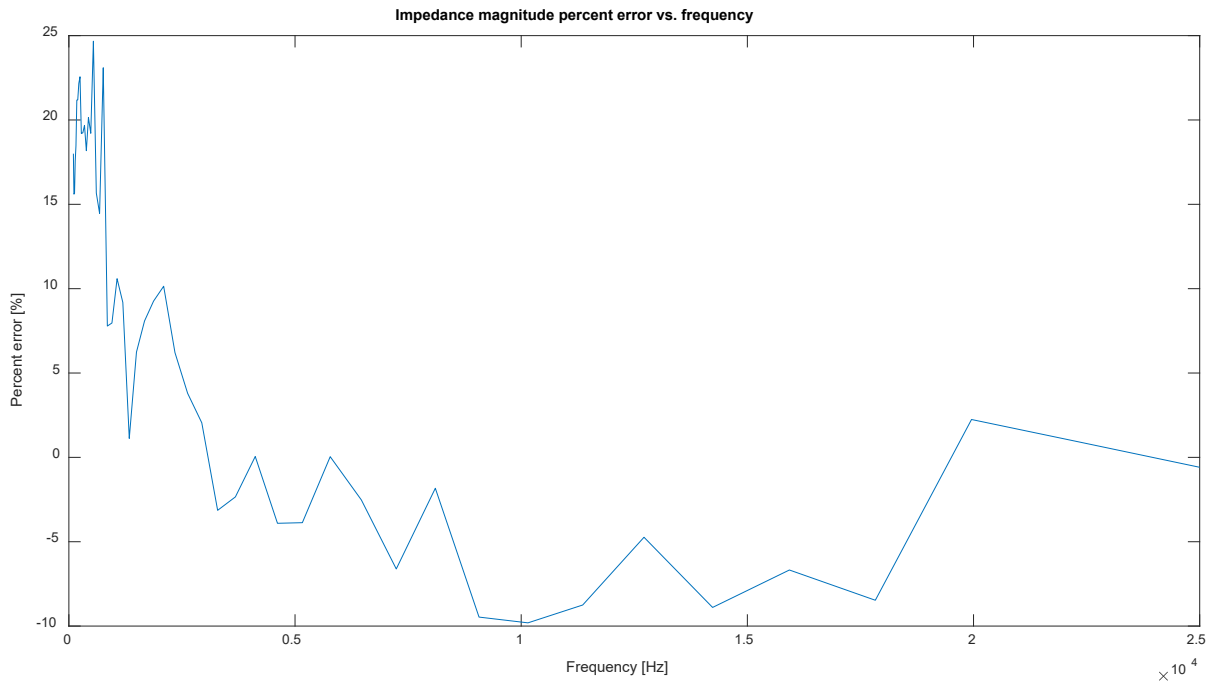

Figure 12: Impedance magnitude percent error vs. frequency

**CV scan test:** The CV scan displays the resulting current for a 10kΩ resistor, with a voltage sweep rate of 1000mV/s. The DAC start voltage is programmed to -1250mV, and the final voltage is 1250mV. However, the actual output voltage from the control amplifier is scaled by a factor of 2.5, so the applied voltage range to the test resistor is from -500mV to 500mV. This same configuration (1000mV/s and -500mV to 500mV) is used for the benchtop potentiostat. By comparing the plots in Figure 13, a time shift between voltage and current samples is apparent—the scan from the developed potentiostat is shifted left compared to the benchtop potentiostat. This time shift is attributed to the ADC sampling and conversion time, which remains constant throughout the scan cycle.

Current is measured using an inverting transimpedance amplifier (TIA), requiring the resulting voltage to be inverted in the developed potentiostat's firmware. The plot shows a linear relationship between voltage and current, as expected. However, the CV scan also reveals that the data from the developed potentiostat contains more noise compared to the data from the benchtop potentiostat.

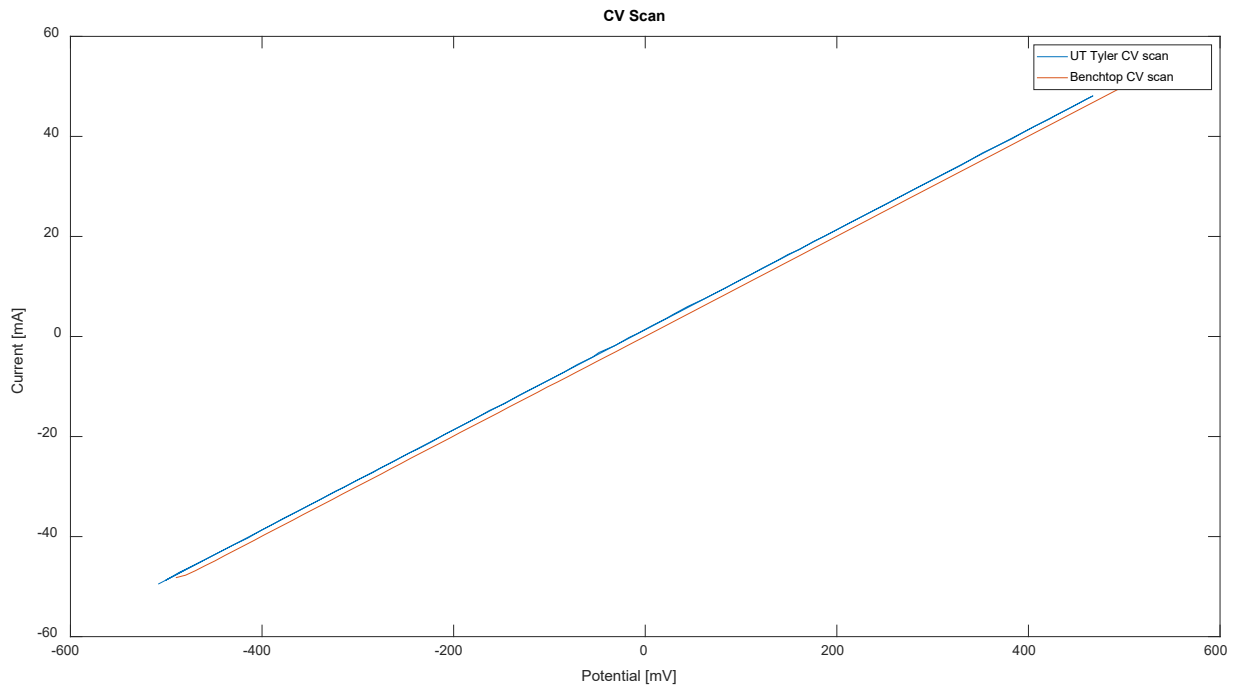

Figure 13: Comparison of CV scans for developed and benchtop potentiostats

**Power tests:** The potentiostat's analog circuits consume 8.2mA of current, with the Bluetooth module in sleep mode. During analog-to-digital conversion and signal processing, the MCU draws 26mA. The total power consumption during measurements is 136.8 mW (34.2 mA at a 4.0V power supply). The firmware is not optimized for low-power usage. Additionally, the analog circuit is turned off after each measurement to conserve energy. However, the prototype potentiostat lacks a power switch to manually turn off the analog circuit, and the current firmware does not support low-power operation or an energy-efficient measurement process.

# Conclusion and Future Scope

Based on existing test data and the availability of miniature electronic components, it is feasible to develop a low-power potentiostat for wearable sensors. However, tests with the prototype potentiostat have highlighted several issues that need attention. Power consumption was not a primary focus during this phase of research and development, which primarily concentrated on data acquisition and processing with a minimalistic approach using simple hardware and signal processing. During electrochemical impedance spectroscopy (EIS) testing, a significant phase error was observed between 150Hz and 250Hz, and from 1.5kHz to 1.6kHz. Although extensively investigated during debugging, the exact cause remains unclear, with possibilities including issues with the ADC hardware or the low-level operations of the ADC driver in the Hardware Abstraction Layer (HAL).

The prototype uses a single ADC with multiplexed inputs, meaning current and voltage are not measured simultaneously during impedance scans. This delay is dynamically compensated by converting it into a phase offset that is subtracted from the measured phase shift. A timing glitch in the microcontroller may explain the phase error at certain frequencies. During cyclic voltammetry (CV) scans, a similar delay exists between current and voltage measurements due to the use of a single ADC. This delay, which is 653μs, is caused by the ADC clock speed of 1MHz. A potential solution is to modify the firmware to use two independent ADCs for simultaneous voltage and current measurements, following a process described by ST Microelectronics. This change could resolve the phase measurement issue in impedance scans. Additionally, adding a calibration table with multiple calibration points could improve impedance magnitude measurements, as the current firmware only has offset calibration for the transimpedance amplifier (TIA) and electrometer.

The prototype is designed for integration with wearable sensors, where low power consumption is crucial. Power-saving measures include switching off the analog circuit during idle times using a p-channel MOSFET controlled by the microcontroller's GPIO pin. Further power savings could be achieved by putting the microcontroller and Bluetooth transceiver in sleep mode during idle periods, though this would require firmware updates without needing hardware changes.

The Bluetooth subcircuit has been verified with demo firmware, but Bluetooth functionality for wireless data transfer requires further firmware development, though it is a lower priority for initial testing. To optimize the design for wearable applications, smaller passive components (e.g., 0402 or 0201 size resistors and capacitors) should be used, and a flexible substrate should replace the current FR4 rigid PCB. A stacked multi-board design could also be explored depending on the integration with wearable sensors.

In summary, while the prototype potentiostat is promising, further testing and size optimization are needed to create a functional, wearable potentiostat.

## Author Contributions

F. F. designed the system, conducted data collection, and analyzed the data. S. T. conceived and supervised the work. All authors participated in drafting and revising the manuscript.

## Acknowledgement

S. T. acknowledges the funding support from the National Science Foundation under award number ECCS 2138701 and the VentureWell under award number 21716-20.